\documentclass[a4paper,12pt]{article}

\usepackage{theorem}
\usepackage{amsfonts}
\newfont\fiverm{cmr5}

\usepackage[dvips]{epsfig}	
\input epic.sty
\linethickness{0.4pt}
\def\tr{{\rm tr\,}}
\def\<{\langle}
\def\>{\rangle}

\makeindex

\begin{document}

\vspace*{-2.5cm}

\begin{center}

{\LARGE \bf Collapse challenge}\\

{\LARGE \bf for interpretations} \\

{\LARGE \bf of quantum mechanics} \\

\vspace{1.5cm}

\centerline{\sl {\large \bf Arnold Neumaier}}

 \vspace{0.5cm}

\centerline{\sl Fakult\"at f\"ur Mathematik, Universit\"at Wien}
\centerline{\sl Nordbergstr. 15, A-1090 Wien, Austria}
\centerline{\sl email: Arnold.Neumaier@univie.ac.at}
\centerline{\sl WWW: http://www.mat.univie.ac.at/$\sim$neum/}

\end{center}


{\bf Abstract.} 
The collapse challenge for interpretations of quantum mechanics 
is to build from first principles and your preferred 
interpretation a complete, observer-free quantum model of the
described experiment (involving a photon and two screens), together 
with a formal analysis that completely explains the experimental result.
The challenge is explained in detail, and discussed in the light
of the Copenhagen interpretation and the decoherence setting.

\vfill
\begin{flushleft}
\end{flushleft}

{\bf E-print Archive No.}: quant-ph/yymmdddd
   \\
{\bf 2002\hspace{.4em} PACS Classification}: 
   \\
{\bf 2000\hspace{.4em} MSC Classification}: primary 
secondary


\section{The challenge} 

In spite of 80 years of work (see, e.g., the reprrint collection by
{\sc Wheeler \& Zurek} \cite{WheZ}), the foundations of quantum 
mechanics are still full of riddles, partly due to the vagueness
of the goal to be explained. This paper proposes a simple test
that the ultimate interpretation would have to meet.

\bigskip

A single photon is prepared in a superposition of two beams. 
A photosensitive screen blocks one of the two beams but has a big hole 
where the other beam can pass without significant interference. 
At twice the distance of the first screen, a second photosensitive 
screen without hole is placed. 

The experimental observation is that the photon is observed at exactly 
one of the two screens, at the position where the corresponding beam 
ends (and {\bf not} in a superposition or mixture of these two 
possibilities).

The challenge is to build from first principles and your preferred
interpretation a complete, observer-free quantum model of this 
experiment (one photon, two screens, and an appropriate environment), 
together with a formal analysis that completely explains the 
experimental result.

{\bf Remark.} 
This challenge was first posed (in essentially unaltered form) 
on June 28, 2004 in the newsgroup sci.physics.research.
The following discussion makes the challenge more precise and evaluatues
how the two most frequently invoked interpretations of the measurement 
process fare in the challenge. Useful additions to this discussion 
will be made available on the WWW \cite{CC}.

\section{Comments}

1. The experimental result has the natural interpretation that the 
photon was either stopped by the first screen, or passed that screen 
successfully. This property is essential for the analysis of any 
quantum experiment which uses screens with holes (or similar filters) 
to create or select beams of particles. Thus reproducing this 
experiment correctly is a basic requirement for any interpretation 
claiming to provide complete foundations for quantum mechanics.

2. Note that it is possible (though not easy) to prepare states with 
definite photon number to reasonable accuracy; see
{\sc Varcoe} et al. \cite{VarBWW}.
Combining this with a half-silvered mirror is a way to achieve the 
preparation required in the challenge.

3. Unitary dynamics demands that the system (photon,screen1,screen2),
characterized -- after tracing out all other degrees of freedom --
by basis states of the form
\[
   |\mbox{photon number, first screen count, second screen count}\>,
\]
evolves from a pure initial state $|1,0,0\>$ into a superposition 
of $|0,1,0\>$ and $|0,0,1\>$, while agreement with experiment demands 
that the final state is either $|0,1,0\>$ or $|0,0,1\>$. 
This disagreement is the measurement problem in its most basic form.

4. Clearly, the experimental result is something completely objective, 
about which all competent observers agree. (This is my definition of 
objectivity.) Thus the analysis is not permitted to
have any dependence on hypothetical observers. 

5. Memory, records, etc. are permitted only if they are modelled as 
quantum objects, too, and the properties assumed about them (such as 
permanence or copyability) are derived from first principles, too.

6. Position, momentum, and time are required to be modelled explicitly;
apart from that, appropriate simplifications are permitted.
For example, it is ok to treat the photon as a scalar particle, 
to restrict to a single space dimension, or to choose a tractable 
interaction. 

7. Approximations are allowed to make the mathematics more tractable; 
but approximations that require for their justification a collapse 
argument are forbidden

8. One can calculate observation probabilities by calculating 
interactions of the photon with a single electron in the screen 
(which is emitted and later magnified if the photon is observed), 
which is fine (and explains everything if the collapse is assumed). 
But this does not help to solve the collapse problem itself. 
Calculating S-matrix elements only means that one then knows the 
superposition into which a state develops; but the challenge is 
about how this superposition of the possible outcomes with their 
associated probabilities collapses into one of the observed states. 
Why does one not end up in a superposition of the state where 
an electron is emitted (and observed by macroscopic magnification) 
from the first screen only, and the state where an electron 
is emitted (and observed by macroscopic magnification)
from the second screen only? Such macroscopic superpositions are 
not observed. 

9. A subjective probabilistic answer is not satisfactory since we have 
only a single photon. What makes different physical observers agree 
that the first screen and not the second detected the photon? 
Clearly, this question is within the realm of physics and should be 
answerable by a fundamental theory underlying all of physics.

\section{Analysis of the Copenhagen interpretation}

The Copenhagen interpretation (see, e.g., 
{\sc von Neumann} \cite{vNeu}) -- which renounces unitarity at 
measurements -- is unsurpassed in its simplicity, and
almost meets the challenge.

Indeed, in the Copenhagen interpretation,
the state remains $|1,0,0\>$ until the photon feels the presence 
of the first screen. In the next split moment, the state collapses,
due to interaction with the classical screen, to either $|0,1,0\>$ or 
$|1,0,0\>$. In the first case, the photon is destroyed and we reached 
a stationary state. In the second case, the state remains $|1,0,0\>$  
until the photon feels the presence of the second screen. 
In the next split moment, the state turns into $|0,0,1\>$, due to 
interaction with the second screen. 

The only thing missing is the required quantum model of the screens.
Although very successful in all situations where the experimental 
setting can be interpreted classically, this unresolved 
quantum-classical interface issue (including the missing definition 
of which situations constitute a measurement) is a serious defect 
of the Copenhagen interpretation when viewed as a fundamental 
interpretation of quantum mechanics.

\section{Analysis of decoherence interpretations}

Decoherence scenarios (see, e.g., {\sc Joos} et al. \cite{JooZK}) 
go something like the following.

Localized position states (of the emitted electron; the photon is 
absorbed hence no longer exists) are robustly selected by screen 
plus environment. This can be justified roughly by noting that the 
interaction between the photon and the emitted electron is given by 
some local operator.
(While some handwaving is involved in this argument since there are
many electrons but only one is emitted; this can probably be cured 
by an appeal to quantum field theory.)

Assuming this robustness and some handwaving that can be made more
precise, decoherence shows that the reduced state of 
(photon + two screens) is not a superposition but a mixture of the two 
states $\phi = |0,0,1\>$ and $\chi = |0,0,1\>$ in question. 
The reduced state is $\rho = (\phi \phi^* + \chi \chi^*)/2$, 
if initially both beams had the same intensity.

But this mixture cannot be interpreted as an ensemble of pure states
in one of the two robust configuraions since it is the partial trace
of a pure state, and hence something irreducible. It is not allowed 
to treat $\rho = \tr_E \psi \psi^*$ (where $\psi$ is the state of 
photon+screens+environment, and the trace is over the environment $E$) 
as an ensemble consisting of 50\% copies of $\phi \phi^*$ and 
50\% copies of $\chi \chi^*$ -- not even for a large stream of 
photons -- since there is no way to decompose $\psi \psi^*$ into two 
states whose partial traces are $\phi \phi^*$ and $\chi \chi^*$. 
But in fact we only have a single photon, 
and there it is completely ridiculous. What one actually observes is 
one of $\phi \phi^*$ and $\chi \chi^*$, and not the mixture.

Erich Joos, one of the exponents of decoherence theory and coauthor 
of the book \cite{JooZK} on decoherence, explicitly
states this missing step in the last paragraph of p.3 in
\cite{Joo1}. The same conclusion is reached in 
the excellent article by {\sc Schlosshauer} \cite{Schlo}.

Thus decoherence only fakes the real situation, and does not explain 
the collapse.

\bigskip

\end{document}